# Mitigating Clipping Effects on Error Floors under Belief Propagation Decoding of Polar Codes

Ahmed Elkelesh, Sebastian Cammerer, Moustafa Ebada and Stephan ten Brink
Institute of Telecommunications, Pfaffenwaldring 47, University of Stuttgart, 70569 Stuttgart, Germany
{elkelesh,cammerer,ebada,tenbrink}@inue.uni-stuttgart.de

*Abstract*—In this work, we show that polar belief propagation (BP) decoding exhibits an error floor behavior which is caused by clipping of the log-likelihood ratios (LLR). The error floor becomes more pronounced for clipping to smaller LLR-values. We introduce a single-value measure quantifying a "relative error floor", showing, by exhaustive simulations for different lengths, that the error floor is mainly caused by inadequate clipping values. We propose four modifications to the conventional BP decoding algorithm to mitigate this error floor behavior, demonstrating that the error floor is a decoder property, and not a code property. The results agree with the fact that polar codes are theoretically proven to not suffer from error floors. Finally, we show that another cause of error floors can be an improper selection of frozen bit positions.

## I. INTRODUCTION

Polar codes were introduced by Erdal Arıkan [1] as a new family of error-correcting codes that provably achieve the symmetric capacity of any symmetric Binary Input Discrete Memoryless Channel (BI-DMC) under successive cancellation (SC) decoding. However, the sub-optimal performance of finite length polar codes under SC decoding has raised the need for more advanced, and probably more complex decoding algorithms. Polar codes were shown to outperform state-of-the-art coding schemes when decoded under successive cancellation list (SCL) decoding [2]. However, the high complexity under SCL decoding and its inherently serial decoding algorithm can be seen as one main drawback of polar codes when compared to low density parity check (LDPC) codes. As an alternative, an iterative algorithm based on the idea of message passing over the encoding graph has been proposed [3] and intensively studied in different regimes. Although belief propagation (BP) decoding of polar codes is outperformed by state-of-the-art SCL decoding in terms of bit error rate (BER) performance, improving its performance is an active area of research [4][5][6] due to the high potential of parallel implementations [7][8], which makes it practical for high-speed applications, and the possibility of soft-in/soft-out decoding, i.e., for joint detection and decoding.

Strongly connected with iterative decoding schemes comes the study of the so-called error floor regime of iterative decoding, as many recent communication applications and storage systems require a rather low BER. The error floor analysis is a well-established area of research extensively explored for different iterative coding schemes such as turbo codes and LDPC codes. For the case of turbo codes, the low-weight codewords have been pointed out to dominate the error floor behavior of the code [9]. One of several ways to increase the minimum distance of the code, i.e., avoid most of the low-weight codewords, is to concatenate the code with an outer error-correcting code such as a high-rate Cyclic Redundancy Check (CRC) code or a Bose-Chaudhuri-Hocquenghem (BCH) code. For the case of a Binary Erasure Channel (BEC) with erasure probability $\epsilon$, the error floor performance of an LDPC code is completely dominated by the (small) stopping sets in the Tanner graph of the code [10]. For the additive white Gaussian noise channel (AWGNC), a more general term of the *stopping sets* is denoted by *trapping sets* (or near codewords) in [11]. The iterative decoding algorithm gets stuck in such error patterns that are inherent to the code structure, no matter how good the channel quality becomes. It is worth mentioning that in some LDPC code constructions, the error floor is caused mainly due to low minimum distance.

Although the log-likelihood ratios (LLR) range is theoretically unbounded, in practice it has to be *clipped* to avoid numerical instability of the boxplus function [12] and due to hardware constraints (i.e., higher complexity due to the need of wider bit widths to represent the messages). The effect of LLR-clipping on the error floor behavior has been extensively studied for LDPC codes, [13][14][15]. In [13], the effect of LLR-clipping on the error floor is studied and proved to be a dominant effect in case of AWGNC. In [14], the upper bound on the error probability was proved to be arbitrarily small as the value of the LLR-clipping $LLR_{max}$ grows large enough, and hence lowering the saturation level of the LLRs. The same conclusion was pointed out in [15], where the trade-off between the numerical stability and the LLR-clipping was discussed.

For the family of polar codes, in [16] it is analytically shown that the size of the minimal stopping sets of BP decoding of polar codes scales with $O(\sqrt{N})$ which reflects a desirable error floor performance. They provide simulation-based observations of the absence of an error floor down to BERs of $10^{-9}$ and $10^{-12}$ for AWGNC and BEC, respectively. Besides that, the authors in [17] proved the absence of error floors for polar codes by considering the scaling of the error probability $P_e$ with the Bhattacharyya parameter $Z(W)$.

However, several results in some recent work indicate the existence of error floors for polar codes under BP decoding, e.g., [18, Fig. 4][19, Fig. 5][20, Fig. 5][21, Fig. 6][7, Fig.

This work has been supported by DFG, Germany, under grant BR 3205/5-1.

5]. The objective of this work is to help resolve the issue of error floor existence in iterative polar decoding and discuss the reasons why such behavior might exist or not. Furthermore, BP decoding of polar codes heavily relies on the usage of the boxplus function, which motivated this work, to better understand the effect of LLR-clipping with respect to error floor performance of polar codes. Note that, one BP decoding iteration for polar codes is based on a possibly large number of successive boxplus operations, potentially leading to clipping issues at much lower number of iterations, particularly when compared to one BP decoding iteration for LDPC codes.

## II. POLAR CODES AND ITERATIVE DECODING

### A. Polar Codes

Polar codes introduced by Arıkan in [1] are the first type of channel codes that are theoretically proven to achieve the channel capacity under a low complexity SC decoding for infinite code lengths. They are based on the theoretical concept of channel polarization, in which $N$ identical channels are combined based on a $2 \times 2$ kernel to produce $N$ synthesized channels showing a polarization behavior.

Channel polarization means that the synthesized channels converge towards either a pure noiseless channel or a pure noisy channel. This polarization effect becomes more obvious when the number of channels to be combined $N$ approaches infinity. Based on this concept, uncoded information bits are transmitted over the noiseless channels and a known sequence of frozen bits are transmitted over the noisy channels. Afterwards comes the polar code construction, which is the phase of selecting the set of $k$ synthesized bit channels upon which uncoded information bits are transmitted. The output of this phase is usually denoted as the information set $\mathbb{A}$. Throughout this paper (except random $\mathbb{A}$ in the last section), we use the polar code construction based on Arıkan's Bhattacharyya bounds [1] of bit channels designed at $Es/N_0 = 0\,\mathrm{dB}$. Finally, the polar encoding is based on the kernel used, which is the basic building block of the encoder. The generator matrix of polar codes of block size $N = 2^n$ is given by

$$\mathbf{G}_N = \mathbf{F}^{\otimes n}, \qquad \mathbf{F} = \begin{bmatrix} 1 & 0 \\ 1 & 1 \end{bmatrix} \qquad (1)$$

where $\mathbf{F}^{\otimes n}$ denotes the $n$th Kronecker power of $\mathbf{F}$. The codewords can be now obtained by $\mathbf{x} = \mathbf{u}\mathbf{G}_N$, where $\mathbf{u}$ contains $k$ information bits and $N-k$ frozen bits.

The performance of polar codes in the finite length regime depends mainly on the type of decoder used. There are two main polar decoding algorithms, the SC decoding (and its variants, e.g., SCL) and the BP decoding.

### B. Belief Propagation Decoding

Belief propagation decoding of polar codes was introduced in [3] to enhance the BER performance of finite length polar codes. The algorithm is based on Gallager's BP decoding for LDPC codes, which is a message passing algorithm in which the information bits are retrieved through iterations.

Two types of messages are involved, left to right messages (**R**-messages) and right to left messages (**L**-messages). The **R**-messages at stage 1 represents the a priori information available to the decoder. The **R**-messages at stage 1 are either 0 or $\infty$[1] for non-frozen and frozen bits, respectively:

$$R_{i,1} = \begin{cases} \infty \text{ (or } LLR_{max}) & i \in \bar{\mathbb{A}} \\ 0 & \text{otherwise.} \end{cases} \qquad (2)$$

The **L**-messages at stage $n+1$ carries the LLR channel output $L_{ch,i}$:

$$L_{i,n+1} = L_{ch,i}. \qquad (3)$$

The **L**- and **R**-messages iteratively propagate through the polar factor graph as shown in Fig. 1. The polar factor graph consists of $\log_2(N) \cdot \frac{N}{2}$ processing element (PE)s. A single PE is shown in Fig. 2. The **L**- and **R**-messages are updated in each PE as follows:

$$\begin{aligned} R_{\text{out},1} &= f(R_{\text{in},1}, L_{\text{in},2} + R_{\text{in},2}) \\ R_{\text{out},2} &= f(R_{\text{in},1}, L_{\text{in},1}) + R_{\text{in},2} \\ L_{\text{out},1} &= f(L_{\text{in},1}, L_{\text{in},2} + R_{\text{in},2}) \\ L_{\text{out},2} &= f(R_{\text{in},1}, L_{\text{in},1}) + L_{\text{in},2} \end{aligned}$$

where $f(L_1, L_2) = L_1 \boxplus L_2$ is commonly referred to as *boxplus* operator [12]. For numerical stability the boxplus function can be reformulated as

$$\begin{aligned} f(x, y) = x \boxplus y &= \ln \frac{1 + e^{x+y}}{e^x + e^y} \\ &= \text{sign}(x) \cdot \text{sign}(y) \cdot \min(|x|, |y|) \\ &\quad + \ln\left(1 + e^{-|x+y|}\right) - \ln\left(1 + e^{-|x-y|}\right). \end{aligned} \qquad (4)$$

Neglecting the two ln-terms in equation (4) leads to the min-approximation. This approximation avoids the (potentially unstable) division, i.e., provides better numerical properties than the original boxplus function, without performance degradation, and thus is more suitable for decoder implementations.

The conventional BP decoder terminates when a pre-defined maximum number of iterations is reached. However, early stopping conditions can be used to speed up the decoding process [22]. Finally, a hard decision is applied to recover the information bits (in the vector $\mathbf{u}$) and the transmitted codeword $\mathbf{x}$ as follows:

$$\begin{aligned} \hat{u}_i &= \frac{1}{2} \cdot (\text{sign}(L_{i,1} + R_{i,1}) + 1) \\ \hat{x}_i &= \frac{1}{2} \cdot (\text{sign}(L_{i,n+1} + R_{i,n+1}) + 1). \end{aligned}$$

---

[1]One could assign the maximum LLR-clipping value ($LLR_{max}$) to the frozen positions.

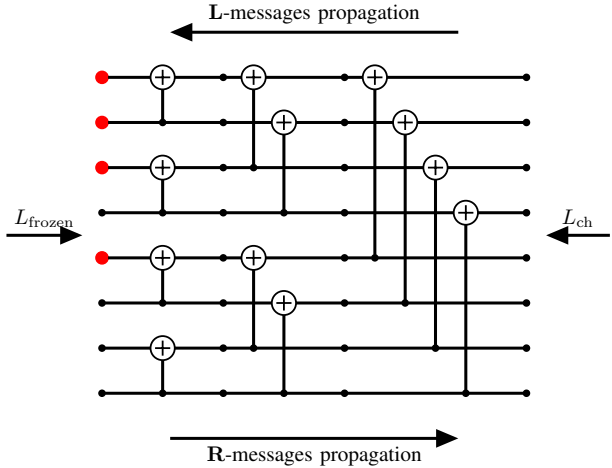

Fig. 1: $N = 8$ polar code factor graph.

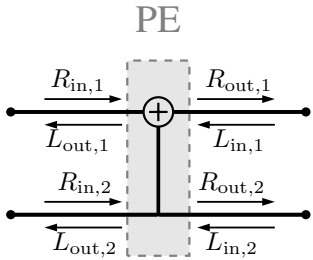

Fig. 2: Processing Element.

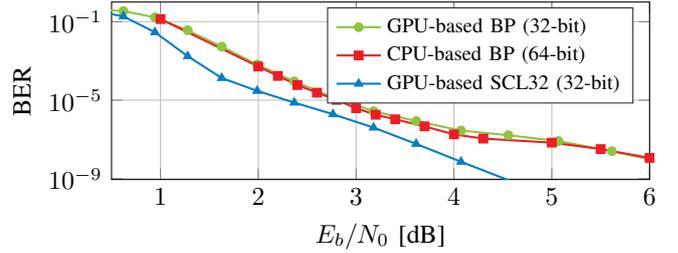

Fig. 3: BER performance of an $N = 4096$, $R = 0.5$ polar code, under BP decoding with clipping value $LLR_{\max} = 20$ for 32-bit and 64-bit numerical precision, and under SCL decoding with list size $L = 32$.

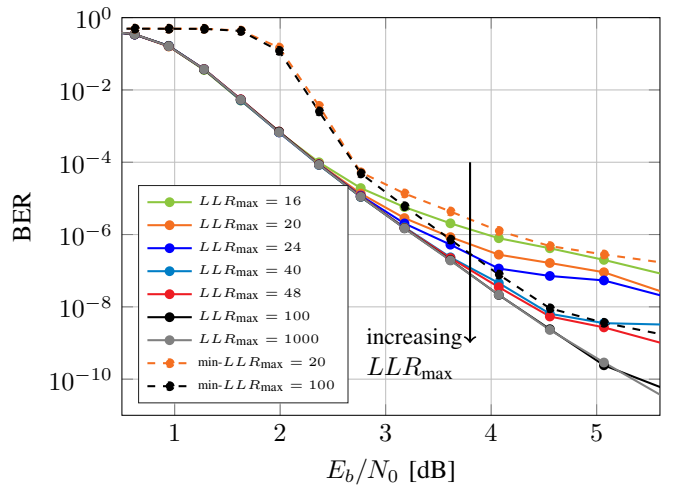

Fig. 4: BER performance of an $N = 4096$, $R = 0.5$ polar code under boxplus-based BP decoding and under *min-approximation*-based BP decoding with different clipping values $LLR_{\max}$.

## III. OBSERVATIONS

Throughout this work, the graphical processing unit (GPU)-based environment of [8] is used for the error rate simulations. As the GPU system is optimized for 32-bit precision (*float*), all simulations are performed with 32-bit operations. Fig. 3 also shows 64-bit precision results (*double*) in order to verify that the behavior is, basically, the same. In the remaining part of the paper we restrict ourself to 32-bit precision. Fig. 4 shows the BER performance while using various clipping values. Additionally, Fig. 5 provides the corresponding block error rate (BLER). Obviously, the error floor vanishes when increasing the clipping value.

As it is, in general, difficult to quantify the error floor, we propose a single-value measure of the *relative* error floor similar to [23], denoted as normalized error (NE)

$$\mathrm{NE}(LLR_{\max}) = \frac{1}{N_s} \sum_{s=1}^{N_s} \frac{\mathrm{BER}_{LLR_{\max}}(\rho_s)}{\mathrm{BER}_{LLR_{\max,\mathrm{ref}}}(\rho_s)}. \quad (5)$$

The NE evaluates the BER at several signal-to-noise-ratio (SNR) points and relates them to a reference curve. In our case the reference curve uses a clipping of $LLR_{\max,\mathrm{ref}} = 100$. Let $\rho_s$ denote the SNR (measured as $E_b/N_0$), and let $\mathrm{BER}_{LLR_{\max}}(\rho_s)$ be the BER achieved at $\rho_s$ for a clipping with $LLR_{\max}$. Obviously, for NE = 1, the error floor achieves the same behavior as the reference curve. In the sequel, we compute the NE over $N_s = 10$ different SNR points from $0.5\,\mathrm{dB}$ to $4.5\,\mathrm{dB}$ with $5 \cdot 10^8$ codewords for each SNR point.

For the NE evaluation, we use the same information set $\mathbb{A}$ and the same parameters in each simulation; thus, the BER only differs in the error floor region as it can be seen in Fig. 4.

We observe in Fig. 4 and Fig. 6 that the NE decreases with $LLR_{\max}$. The NE for different block lengths is also depicted in Fig. 7. The NE increases with $N$, as the error floor becomes more obvious for long polar codes.

We observe that only a few BP iterations may be needed to fall into a trapping set, such that the bounded LLR-clipping value is reached by the growing LLR-messages. Thus, increasing the maximum number of iterations does not help [13], which means that other, modified BP decoding algorithms will be needed to solve this issue.

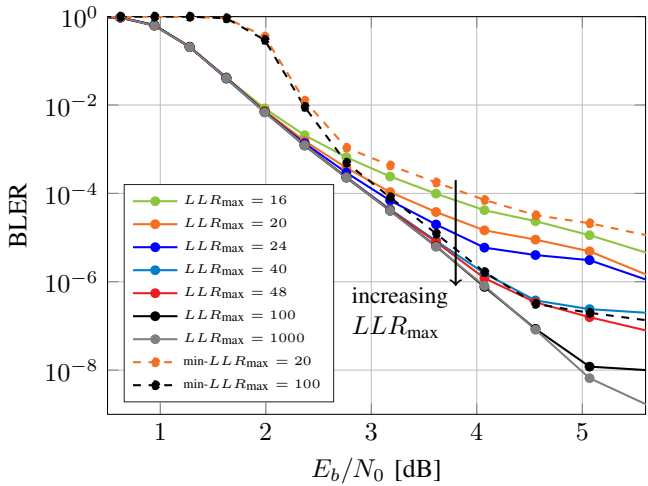

Fig. 5: BLER performance of an $N = 4096$, $R = 0.5$ polar code under boxplus-based BP decoding and under *min-approximation*-based BP decoding with different clipping values $LLR_\text{max}$.

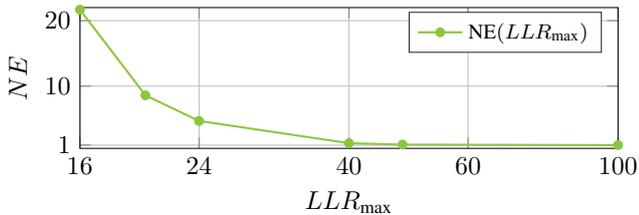

Fig. 6: NE evaluation of an $N = 4096$, $R = 0.5$ polar code under BP decoding with different clipping values $LLR_\text{max}$ and $LLR_\text{max,ref} = 100$.

## IV. ERROR FLOOR CAUSED BY LLR-VALUE CLIPPING

To gain further insight into error floor effects, different decoder modifications mitigating error floors are investigated. The fact that different algorithms can overcome the error floor let us conclude that the information itself is not destroyed, i.e., the observed error floor is a *decoder property* and not a *code property* (as it might happen for a weak distance spectrum), similar to [11]. This can also be seen in Fig. 3, as the SCL decoder shows no error floor behavior (and also the BP decoder for higher LLR-clipping values tends to show no error floor), i.e., the code itself can be decoded without error floor.

In the following, we propose four different modifications to the conventional BP decoding algorithm to overcome the error floor caused by small LLR-clipping values, yet, at the expense of higher decoding complexity. The idea is to try to achieve the same performance as a BP decoder with a very high $LLR_\text{max}$ (e.g., 100) using a modified BP decoder with a low $LLR_\text{max}$ (e.g., 20), to keep implementation complexity small. Thus, in the following, we propose some algorithms to be combined with the BP decoder with a low $LLR_{max}$.

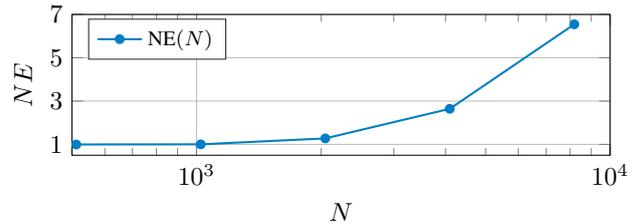

Fig. 7: NE evaluation for different $N$ of a $R = 0.5$ polar code under BP decoding with clipping values $LLR_\text{max} = 20$ and $LLR_\text{max,ref} = 100$.

Test sets were collected at $E_b/N_0 = 5\,\text{dB}$. The test sets, in this context, are the scenarios in which the BP decoder converges to the correct codeword using a high $LLR_{max} = 100$, while it fails to converge using a low $LLR_{max} = 20$. The *success rate* for each algorithm modification is shown in Tab. I.

### A. Guessing Algorithm

In the high $E_b/N_0$ (low BER) region, if the BP decoder fails to converge, it is observed that there are usually very few number of bits in error due to short cycles. Knowing the correct value of one (up to three) of these bits helps the BP decoder to converge and escape this small-sized trapping set.

In scenarios that did not converge, the presence of oscillating LLR-signs of some bits was observed. Thus, the idea is to guess the sign of one of these oscillating bits and, then, to assign the maximum LLR-value (i.e., $LLR_{max}$ or $-LLR_{max}$) and continue the BP decoding as proposed in [24] for LDPC codes, and in [16] for polar codes. This may help to push the decoder to convergence (i.e., "belief pushing"), and, if not, the guess for the chosen bit is reversed, or apply the same guessing algorithm to another bit, and so on.

### B. Adding Virtual Noise

As guessing either requires a genie-aided decoding or results in a complexity overhead due to multiple guesses, we investigate a more practical approach.

When the BP decoder fails to converge, random virtual noise can be added to the input from the channel, as in [25], according to Fig. 8, i.e.,

$$\tilde{y} = y + n_v, \quad n_v \sim \mathcal{N}(0, \sigma_v^2)$$

In most of the investigated cases in Tab. I, the decoder then can successfully decode $\tilde{y}$, although the effective channel becomes worse.

### C. Modified Boxplus function

Scale the boxplus function output by $\alpha$ (i.e., $0 < \alpha < 1$) [22]. This helps in the low BER region as it prevents LLRs from running into saturation during the decoding iterations and thus, mitigates the error floor.

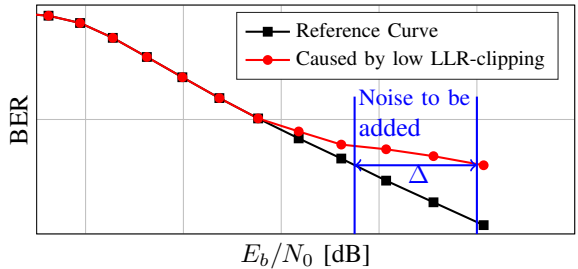

Fig. 8: Visualization of maximum noise power to be added by algorithm B (*Add Virtual Noise*).

TABLE I: Success rate of the different algorithms.

|   | Algorithm | Success rate $\tau$ |
|---|---|---|
| A | Guessing algorithm (1 , 2 , 3 bits) | 0.9585 , 0.9852 , 0.9911 |
| B | Add Virtual Noise $\sigma_v^2 = 0.36$ | 0.9925 |
| C | Modified Boxplus function $\alpha = 0.9375$ | 0.8889 |
| D | Multi-trellis BP decoder | 1 |

*D. Multi-trellis BP decoder*

When the BP decoder that is based on the conventional factor graph fails to converge, a different realization (representation) of the polar factor graph can be used, following the basic concept in [26]. This algorithm helps the overall decoder in avoiding, or, escaping to fall into a trapping set (i.e., due to the different loops in the different realizations of the factor graph). However, we may require a higher number of iterations, i.e., the decoding complexity slightly increases.

## V. OTHER REASONS FOR ERROR FLOORS

Although it is theoretically proven that polar codes do not suffer from error floors [17], the specific *decoder* itself can show such a behavior. In general, avoiding loops and short cycles in iterative decoding is crucial. In order to illustrate such effect, Fig. 9 shows the BER with $m$ randomly selected, additional frozen bit positions. Starting with an information set $\mathbb{A}$ ($R = 0.5$), no error floor can be seen. However, with the new information set $\mathbb{A}_m$ ($R_m = \frac{R \cdot N - m}{N} = R - \frac{m}{N} < R$), which is the same as $\mathbb{A}$ but with extra $m$ frozen *known* bits, an error floor appears. Obviously, the maximum a posteriori (MAP) performance of $\mathbb{A}_m$ must be better than that of $\mathbb{A}$, as we effectively only remove $2^m$ possible codewords. All other codewords are the same[2]. As a result of this example, one can not state that polar codes under iterative decoding do not show an error floor. However, when carefully designed (i.e., code construction), properly implemented and when an appropriate decoder is used, then the error floor indeed vanishes.

## VI. CONCLUSION

In this paper, the effect of the LLR-clipping on error floors of polar codes is studied. Increasing the LLR-clipping values leads to an error floor-free BP decoder. For implementation simplicity, low LLR-clipping values are favorable; thus we proposed four different modifications to the BP decoding algorithm that can overcome error floors while keeping LLR-clipping values small. Thus, the observed error floor is shown to be a *decoder property* and not a *code property*. Further, we identified that, besides clipping, improper polar code construction is yet another reason for error floors in BP decoding as it can result in unfavorable loops in the decoding graph.

[2]We do not state that randomly selecting extra frozen positions is a good approach for improving the code performance. However, the MAP performance can never degrade by such an extension.

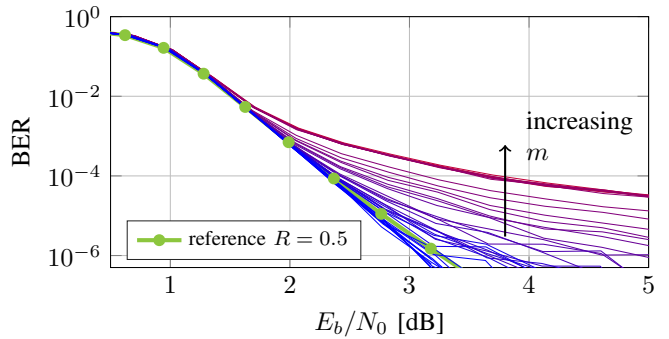

Fig. 9: BER performance of an $N = 4096$ polar code under BP decoding with LLR-clipping value $LLR_{\max} = 100$ for additional frozen bits ($10 \leq m \leq 48$), starting from a rate $R = 0.5$ code.